\pdfoutput = 1
\documentclass{article}

\usepackage{setspace}
\usepackage{arxiv}
\usepackage{indentfirst}
\usepackage{parskip}
\usepackage[utf8]{inputenc} % allow utf-8 input
\usepackage[T1]{fontenc}    % use 8-bit T1 fonts
\usepackage{hyperref}       % hyperlinks
\usepackage{url}            % simple URL typesetting
\usepackage{caption}
\usepackage{bm}
\usepackage{geometry}
\usepackage{booktabs}       % professional-quality tables
\usepackage{amsfonts}       % blackboard math symbols
\usepackage{nicefrac}       % compact symbols for 1/2, etc.
\usepackage{microtype}      % microtypography
\usepackage{lipsum}
\usepackage{amssymb,amsmath,bm}
\usepackage{algorithmicx,algorithm}
\usepackage{algpseudocode}
\usepackage{graphicx}
\usepackage{amsbsy}
\usepackage{bm}
\usepackage{caption}
\usepackage{enumitem}
\usepackage{longtable}
\algrenewtext{EndFunction}{\algorithmicend}
\algrenewtext{EndFor}{\algorithmicend}
\algnewcommand\algorithmicto{\textbf{to}}
\algrenewtext{EndProcedure}{\algorithmicend}
\title{DTF: Deep Tensor Factorization for Predicting
Anticancer Drug Synergy}

\makeatletter
\newenvironment{breakablealgorithm}
  {% \begin{breakablealgorithm}
   \begin{center}
     \refstepcounter{algorithm}% New algorithm
     \hrule height.8pt depth0pt \kern2pt% \@fs@pre for \@fs@ruled
     \renewcommand{\caption}[2][\relax]{% Make a new \caption
       {\raggedright\textbf{\ALG@name~\thealgorithm} ##2\par}%
       \ifx\relax##1\relax % #1 is \relax
         \addcontentsline{loa}{algorithm}{\protect\numberline{\thealgorithm}##2}%
       \else % #1 is not \relax
         \addcontentsline{loa}{algorithm}{\protect\numberline{\thealgorithm}##1}%
       \fi
       \kern2pt\hrule\kern2pt
     }
  }{% \end{breakablealgorithm}
     \kern2pt\hrule\relax% \@fs@post for \@fs@ruled
   \end{center}
  }
\makeatother

\author{
Zexuan Sun\\
School of Mathmatics and Statistics \\
Wuhan University \\
Wuhan, 430072, China\\
   \And
 Shujun Huang \\
  College of Pharmacy\\
  University of Manitoba\\
  Winnipeg, Manitoba, R3E 0T5, Canada \\
  \And
  Peiran Jiang \\ 
  Department of Bioinformatics  \& Systems Biology\\
  Huazhong University of Science and Technology\\
  Wuhan, 430074, China \\
  \And 
  Pingzhao Hu\thanks{Corresponding Author} \\
  Research Institute in Oncology and Hematology \\
  CancerCare Manitoba \\ 
  Winnipeg, R3E 0V9, Canada \\
}

\begin{document}
\maketitle
\begin{abstract}
%\begin{spacing}{1.3}
\textbf{Motivation:}  Combination therapies have been widely used to treat cancers. However, it is cost- and time-consuming to experimentally screen synergistic drug pairs due to the enormous number of possible drug combinations. Thus, computational methods have become an important way to predict and prioritize synergistic drug pairs.

\textbf{Results:} We proposed a Deep Tensor Factorization (DTF) model, which integrated a tensor factorization method and a deep neural network (DNN), to predict drug synergy. The former extracts latent features from drug synergy information while the latter constructs a binary classifier to predict the drug synergy status. Compared to the tensor-based method, the DTF model performed better in predicting drug synergy. The area under precision-recall curve (PR AUC) was 0.57 for DTF and 0.24 for the tensor method. We also compared the DTF model with DeepSynergy and logistic regression models, and found that the DTF outperformed the logistic regression model and achieved  almost the same performance as DeepSynergy using several typical metrics for classification task.  Applying the DTF model to predict missing entries in our drug-cell line tensor, we identified novel synergistic drug combinations for 10 cell lines from the 5 cancer types.  A literature survey showed that some of these predicted drug synergies have been identified in vivo or in vitro. Thus, the DTF model could be a valuable in silico tool for prioritizing novel synergistic drug combinations.

\textbf{Availability:} Source code and data is available at \href{https://github.com/ZexuanSun/DTF-Drug-Synergy}{https://github.com/ZexuanSun/DTF-Drug-Synergy}
%\end{spacing}
\end{abstract}

% keywords can be removed
%\newpage
%\renewcommand{\baselinestretch}{1.5}
\section{Introduction}
Though monotherapy has contributed a lot to helping cure many human diseases, it has several evident drawbacks, such as acquired resistance or low efficiency \cite{1,2}. The complexness of human diseases is usually resulting from the complex interactions of different phenomic and genomic factors. Thus, single drug, which typically targets on a single protein or pathway, is usually hard to treat the complex diseases well. To solve this dilemma, there comes combinatorial drug therapy, which uses a pair of or more drugs simultaneously to treat a specific disease. The synergistic effect of certain drug pairs can potentially improve the curative effect significantly. For instance, pentamidine and chlorpromazine do not exhibit any traces of inhibiting tumor activities while being used individually, however, the combination of these two drugs is able to inhibit the growth of tumor efficiently. What’s more, the drugs used for evaluating drug synergetic effect usually employ existing drugs, which have been studied thoroughly and approved by Food and Drug Administration (FDA) for treating specific diseases. This will save lots of time for clinical trials of the safety of these drug combinations. As a result, drug combination therapies have become a more and more popular treatment option for complex diseases. However, how to identify the drug pairs with drug synergistic effect is still challenging since the search space of the drug pairs from the drugs approved by FDA is huge. It is too time-consuming and unrealistic to implement clinical assays on all drug pairs. Therefore, computational methods for predicting drug pairs with strong synergistic effect are in great demand.

Currently, there are many computational methods for predicting relevant drug pairs. These include both traditional machine learning methods and deep learning methods. For example, Sidorov et al. proposed models for drug synergy prediction based on random forest (RF) and extrEme gradient boosting (XGBoost) \cite{3}. The physicochemical properties of drugs were used as the input of the models. Zhang et al. developed a model, AuDNNsynergy, based on deep learning method, which took advantage of the gene expression, copy number and genetic mutation data coming from cancer cell lines to predict drug pairs with high synergistic effect \cite{4}. 

However, these methods have not made use of structure of the drug synergy data as a multi-way data (e.g. a data set can be represented as a multidimensional array). In fact, multi-way data reflects a structure of multi-way relations, which can be best represented as a multi-way array, that is, so-called tensor \cite{5}. Tensor decomposition methods are utilized to decompose a given tensor constructed from raw data to capture latent re- lations between variables, which can be used for discovering hidden pat- terns or performing classifications. However, classic algorithms to decom- pose tensors cannot handle those tensors with missing values, which is a common issue in predicting drug synergetic effect. Some studies were conducted to solve the problem based on novel tensor frameworks. For example, Chen and Li proposed DrugCom, a tensor-based model, which incorporated multiple different existing data sources related to drugs and diseases. DrugCom decomposed a tensor with missing values by integrating existing knowledge at the same time to get the latent information of drug synergy, and demonstrated a high prediction performance over other methods \cite{6}. Acr \textit{et al.} expanded the most well-known tensor factorization method CANDECOMP/PARAFAC (CP) as a weighted least squares problem that uses a first-order optimization approach to handle the missing values in a given tensor. The new approach was called as CP Weighted OPTimization (CP-WOPT) \cite{5}. This approach can capture the latent structure of the data via a higher-order factorization.

Although tensor-based factorization approach is efficient to represent multi-way data, there is still a much need to improve its prediction performance. Recently, deep learning methods have been shown to predict drug pairs with high synergy scores. Our study tried to combine tensor-based framework and deep learning methods together to predict synergetic effect of drug pairs. We proposed a Deep Tensor Factorization (DTF) model, which is comprised mainly by a tensor factorization method and a deep neural network (DNN). We first use the algorithm, CP-WOPT, to decompose tensor with missing entries, then the results of the tensor decomposition are served as features to be used to train the DNN model, which can predict the synergetic effect of drug pairs. This strategy allows the DTF model to capture the structure of multi-way data and learn more latent information with the help of deep learning method, therefore to enhance its overall performance.

\section{Materials and Methods}

\begin{figure}[htb!]
    \centering
    \includegraphics[width = 9cm]{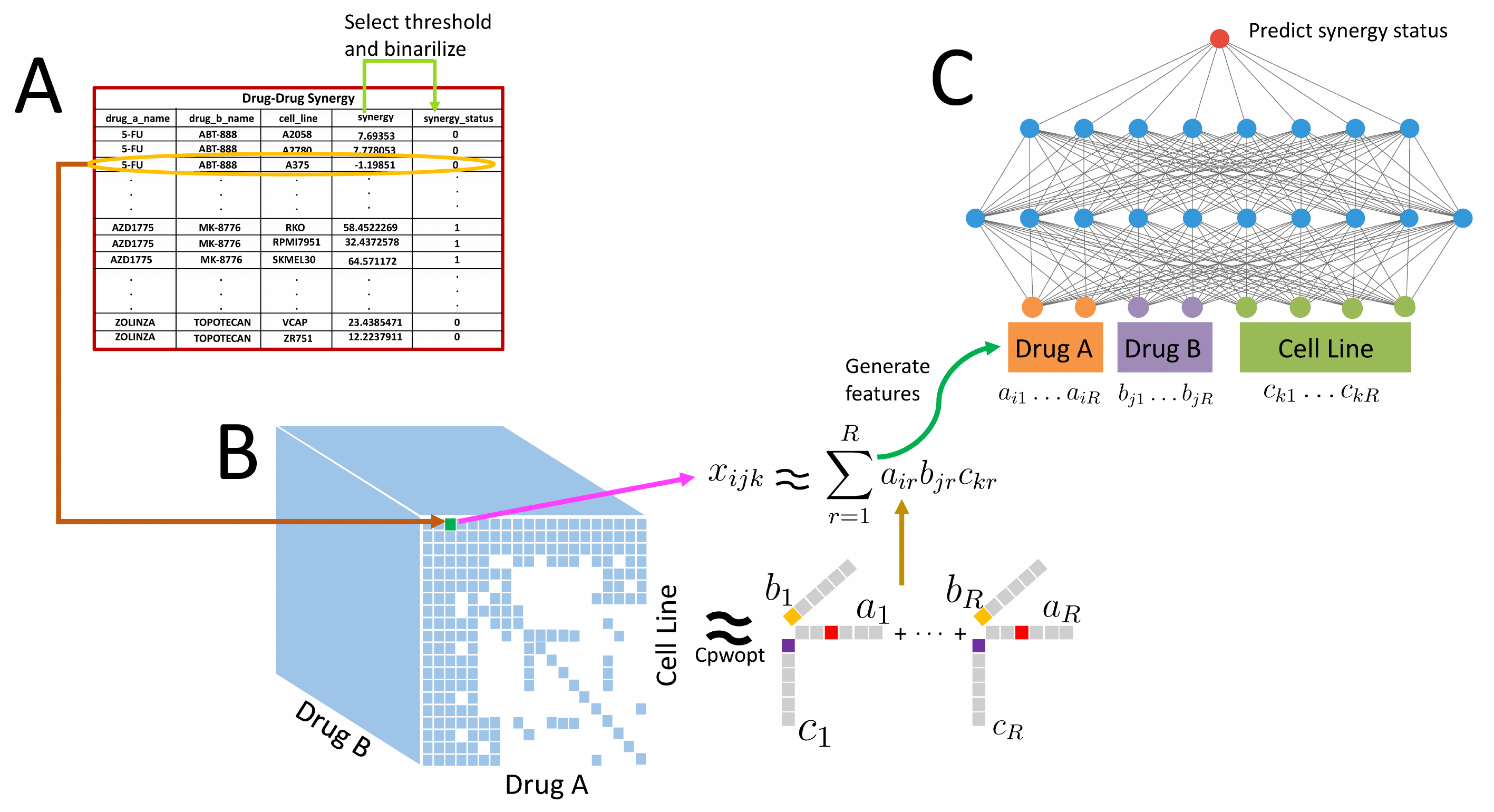}
    \caption{
    \textbf{The workflow of the proposed DTF model}. (A) Data collection and preprocessing. The drug-drug synergy (DDS) data was collected to construct tensor, which was binarized to create synergy status, serving as labels to train the deep neural net- work. (B) Construct the tensor with missing values and decompose it using CP- WOPT. The results of the tensor factorization are used as features to train the deep neural network. (C) Build and train the deep neural network. Use the labels obtained from Step A and the features generated from Step B to train the deep neural network for predicting the synergy status of a given drug pair.
    }
    \label{workflow}
\end{figure}

Our model design is shown in \textbf{Figure} \textbf{\ref{workflow}}. The DTF model for synergistic drug combination prediction is based on two sub-models: a special tensor decomposition model to decompose the tensor with missing values and a DNN model to predict the drug synergy status.

%\newpage

\subsection{Data Collection and Preprocessing}
The data we mainly used is the drug-drug synergy (DDS) data derived from ONeil \textit{et al.}’s study \cite{7}. We got 23,062 drug combinations with the corresponding Loewe synergy scores measured among 38 drugs in 39 cell lines, which come from 6 human cancer types (\textbf{Fig. 1. A}). We also refer the Loewe synergy score as Loewe additivity.

We used the DDS data to construct the tensor. Since we have three variables, i.e., drug A, drug B and cell line, we built a 3-order tensor with three axes representing the three variables, respectively. The value of each entity of the tensor is the synergy score corresponding to a specific drug pair and cell line. We also recorded the positions of the missing values in the tensor for the convenient of tensor decomposition. For some specific cell lines, there were experiments carried out multiple times for the same drug pairs. In order to construct the three-dimensional (3D) drug-drug-cell-line tensor, we averaged these scores for the same drug-drug pairs. It should be also noted that the synergy score of drug A and drug B is the same as that of drug B and drug A. Therefore, for each cell line the matrix formed by the synergy of drug pairs is symmetrical. We set the diagonal of each matrix to zero since there is of course no synergistic effect for the pair of the same drugs.

\begin{figure}[htb]
    \centering
    \includegraphics[width = 9cm]{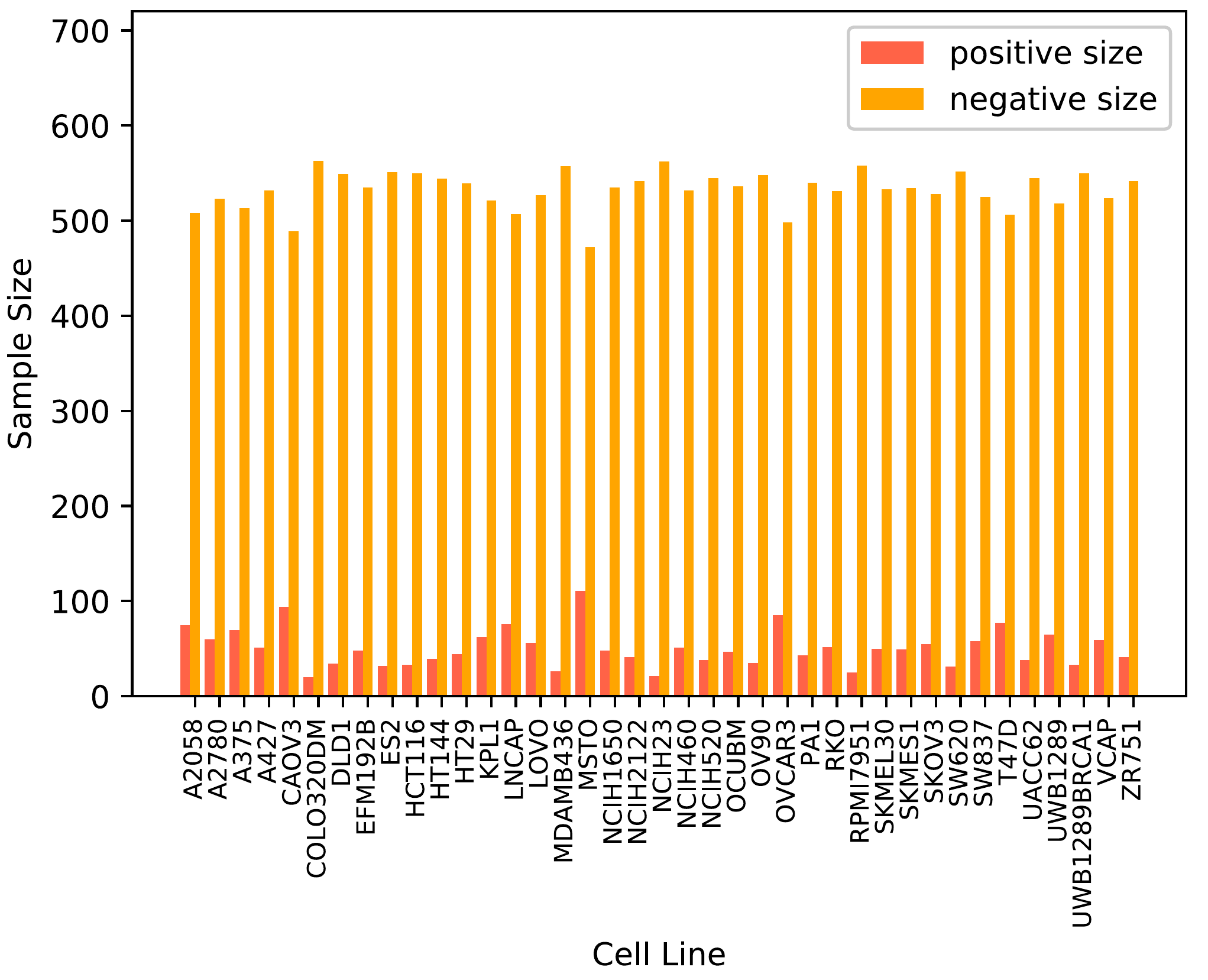}
    \caption{\textbf{Sample size for all cancer cell lines}}
    \label{pos_neg}
\end{figure}

 We also used the DDS data to generate classification labels or synergy status for training the DNN. We select the threshold as 30, which was used in \cite{8}. That is to say, if the synergy score of a given drug pair is greater than 30, the synergy status is 1, otherwise, it is 0. We treated the entities with synergy status 1 as positive samples, and those with synergy status 0 as negative samples. The numbers of positive and negative samples for each cell line are shown in \textbf{Figure \ref{pos_neg}}.

\subsection{DTF: Deep Tensor Factorization}

\subsubsection{Notations}
We define the dimension  of a tensor and multi-way data as \textit{order}, and refer each dimension as a \textit{mode}. We use a 
lowercase letter ($a$) to denote a scalar, a boldface lowercase 
letter ($\bm{a}$ ) to denote a vector,
a boldface capital letter ($\bm{A}$) 
to denote a matrix and a boldface Euler script letter ($\bm{\mathcal{A}}$) 
to denote a tensor. Every element of a tensor is denoted by a lowercase letter with a subscript. For a three order tensor $\bm{\mathcal{X}}$, its entries can be represented as $x_{ijk}$. 

% We define the dimensionality of a tensor and multi-way data as \textit{order}, and refer each dimension as a \textit{mode}. We use a lowercase letter ($a$) to denote a scalar,  a boldface lowercase letter ({$\bm{a}$) to denote a vector, a boldface capital letter ($\textbf{A}$) to denote a matrix and a boldface Euler script letter ($\bm{\mathcal{A}}$) to denote a tensor. Every element of a tensor is denoted by a lowercase letter with a subscript. For a three order tensor $\bm{\mathcal{X}}$, its entries can be represented as $x_{ijk}$.

\textit{Subarrays} (or \textit{subfields}) can be created by fixing some of the given tensor’s indices. If we fix all but one index, \textit{slices} (or \textit{slabs}) are created. If we fix all but two indices, there come fibers. It is easy to find out that for a third order tensor each slice is actually a matrix.

We use $\otimes$ to represent a multi-way vector outer product, which is a tensor and each entry of the tensor is the product of corresponding elements in vectors. For example, the vector outer product of 3 vectors, $\bm{a,b,c}$ is a three-dimensional tensor $\bm{\mathcal{X}}$, where $(\bm{\mathcal{X}})_{i j k}=a_{i} b_{j} c_{k}$. 

Say we have two same-sized tensors $\bm{\mathcal{X}}$ and $\bm{\mathcal{Y}}$, which are of size $I_{1} \times I_{2} \times \cdots \times I_{N}$. We define their Hadamard (elementwise) product as$ \bm{\mathcal{X}} * \bm{\mathcal{Y}}$ where 
$(\bm{\mathcal{X}} * \bm{\mathcal{Y}})_{i_{1} i_{2} \cdots i_{N}}=x_{i_{1} i_{2} \cdots i_{N}} y_{i_{1} i_{2} \cdots i_{N}}$.

For a $I_{1} \times I_{2} \times \cdots \times I_{N}$ sized tensor 
$\bm{\mathcal{X}}$, we define its \textit{norm} as $\|\bm{\mathcal{X}} \|= \sqrt{\langle \bm{\mathcal{X}}, \bm{\mathcal{X}} \rangle}$. Recall that for matrices and vectors, $\|\cdot\|$
can be referred as Frobenius-norm and two-norm, respectively. We are also able to give the definition of a weighted norm. Say 
$\bm{\mathcal{X}}$ and $\bm{\mathcal{W}}$
are two same-sized tensors, then we can define the $\bm{\mathcal{W}}$ weighted norm of 
$\bm{\mathcal{X}}$  as $\|\bm{\mathcal{X}}\|_{\bm{\mathcal{W}}}=\|\bm{\mathcal{W}} * \bm{\mathcal{X}}\|$.

If there are a list of matrices $\bm{A}^{(n)}$
of size$I_{n} \times R$  for $n = 1, \cdots, N$, 
the notation 
$\left[\kern-0.25em\left[
{\bm{A}^{(1)}, \bm{A}^{(2)}, \cdots, \bm{A}^{(N)}}
\right]\kern-0.25em\right]$
gives an $I_{1} \times I_{2} \times \cdots \times I_{N}$ tensor, where
$
(\left[\kern-0.25em\left[
{\bm{A}^{(1)}, \bm{A}^{(2)}, \cdots, \bm{A}^{(N)}}
\right]\kern-0.25em\right])_{i_{1} i_{2} \cdots i_{N}}=\sum_{r=1}^{R} \prod_{n=1}^{N} a_{i_{n r} r}^{(n)}$
, for  $i_{n} \in\left\{1, \ldots, I_{n}\right\}, n \in\{1, \ldots, N\}$.

The rank of a N-way tensor is 1 if an outer product of N vectors equals to this tensor. A N-way tensor is of rank-1 if it can be strictly decomposed into the outer product of N vectors. We define the rank of a tensor $R$ as the mini- mum number of rank-one tensors which are required to get $\bm{\mathcal{X}}$  as their sum.
For instance, a rank-$R$ 3D tensor can, therefore, be written as $\bm{\mathcal{X}} =
\sum_{r=1}^{R} a_{r} \otimes b_{r} \otimes c_{r}
= 
\left[\kern-0.15em\left[
{\bm{A}, \bm{B}, \bm{C}}
\right]\kern-0.15em\right]
$. We refer the matrices $\bm{A}, \bm{B}, \bm{C}$ as factor matrices since they collect vectors from the rank-one components and hold them as columns. It is known that the problem of computing the rank of a tensor is NP-hard. Thus, in practice, we cannot know the exact rank of the tensor we investigate.

\subsubsection{Tensor Decomposition Algorithms}

Rank decomposition is one of the most popular tensor decomposition methods, which stems from the definition of tensor rank. The key idea underlying rank decomposition is to use the sum of a sequence of rank- one tensors to approximate the original tensor. CANonical DECOM- Position (CANDECOMP) and the PARAllel FACtors (PARAFAC) de- compositions are the most popular rank decomposition approaches, which were proposed in different knowledge domains independently. Interestingly, both of them follow similar rules, so we usually name the methods as the CANDE- COMP/PARAFAC or canonical polyadic decomposition (CPD) \cite{9}. For a particular 3D tensor %(\textbf{Figure \ref{cpd}}), the CPD algorithm is to find the optimal $\widehat{\bm{\mathcal{X}}}$ to minimize 
$   \|\bm{\mathcal{X}}-\widehat{\bm{\mathcal{X}}}\|$,   where $\widehat{\boldsymbol{X}} = \sum_{r=1}^{R} a_{r} \otimes b_{r} \otimes c_{r} =
    \left[\kern-0.15em\left[
{\bm{A}, \bm{B}, \bm{C}}
\right]\kern-0.15em\right]$. 
Equivalently, for third-order tensors, the CP decomposition can be treating as optimizing the objective error function as below:

\begin{equation}
f(\bm{A}, \bm{B}, \bm{C})=\frac{1}{2} \sum_{i=1}^{I} \sum_{j=1}^{J} \sum_{k=1}^{K}\left(x_{i j k}-\sum_{i=1}^{R} a_{i r} b_{j r} c_{k r}\right)^{2}
\label{eq}
\end{equation}
More 
details of the algorithm can be found in \cite{9}.

%\vspace*{-7mm}
\begin{figure}[htb!]
    \centering
    \includegraphics[width = 9cm]{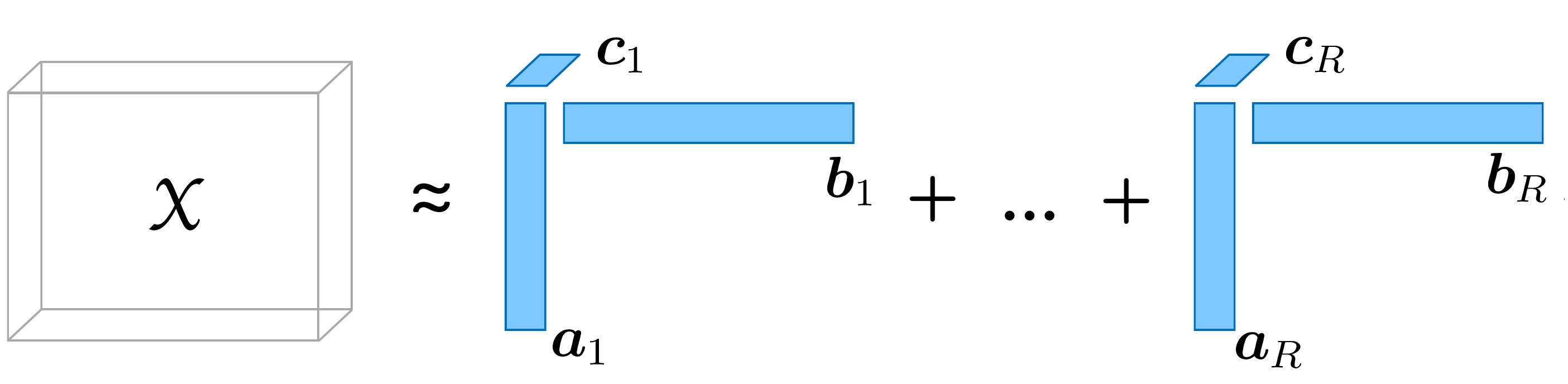}
    %\vspace*{-13mm}
    \caption{\textbf{Illustration of CP decomposition.}}
    \label{cpd}
\end{figure}

% \begin{equation}
%     \min _{\widehat{\bm{\mathcal{X}}}} 
%     \|\bm{\mathcal{X}}-\widehat{\bm{\mathcal{X}}}\|,  \text { where } \widehat{\boldsymbol{X}} = \sum_{r=1}^{R} a_{r} \otimes b_{r} \otimes c_{r} =
%     \left[\kern-0.15em\left[
% {\bm{A}, \bm{B}, \bm{C}}
% \right]\kern-0.15em\right]
% \label{eq1}
% \end{equation}

% \left.\min _{\hat{x}}\|\boldsymbol{x}-\widehat{\boldsymbol{x}}\|, \text { where } \widehat{\boldsymbol{X}}=\sum_{r=1}^{R} a_{r} \otimes b_{r} \otimes c_{r}=\mathbb{I} \bm{A}, \bm{B}, \bm{c}\right]
% \begin{equation*}
%     \min_{\tilde{\bm{\mathcal{X}}}} \| \bm{\mathcal{X}}-\widehat{\bm{\mathcal{X}}} |, \text { where } 
% \end{equation*}

\subsubsection{CP Weighted OPTimization}
The CP decomposition method cannot work When a tensor has missing values. Evrim Acar \textit{et al.} proposed a new mehtod called 
CP Weighted OPTimization (CP-WOPT) to solve this problem. The key of CP-WOPT is to introduce a nonnegtive tensor 
$\bm{\mathcal{W}}$. For a three-dimensional tensor with missing values $\bm{\mathcal{X}}$, we define the $\bm{\mathcal{W}}$, which is of the same size as $\bm{\mathcal{X}}$.
 For each element of $\bm{\mathcal{W}}$, namely, $w_{ijk}$ is 1, if if the corresponding element of $\bm{\mathcal{X}}$, i.e. $x_{ijk}$ is known, otherwise, it is 0. Then we are able to get the weighted version of (\ref{eq}) as follow:
 
 \begin{equation}
\bm{f}_{\bm{\mathcal{W}}}(\bm{A}, \bm{B}, \bm{C})=\frac{1}{2} \sum_{i=1}^{I} \sum_{j=1}^{J} \sum_{k=1}^{K}{w_{ijk}\left(x_{i j k}-\sum_{i=1}^{R} a_{i r} b_{j r} c_{k r}\right)}^{2}
\label{eq2}
\end{equation}
We can easily generalize this to N-way tensor, and get the N-way weighted objective function. Let $\bm{\mathcal{X}}$ be a tensor of size $I_{1} \times I_{2} \times \cdots \times I_{N}$  and assume the rank of $\bm{\mathcal{X}}$ is $R$. Then we Then we can rewrite the objective function using matrices as follow: 
\begin{equation}
\bm{f}_{\bm{\mathcal{W}}}(\bm{A}^{(1)}, \bm{A}^{(2)}, \cdots, \bm{A}^{(N)}) = \frac{1}{2} \| (\bm{\mathcal{X} -
\left[\kern-0.25em\left[
{\bm{A}^{(1)}, \bm{A}^{(2)}, \cdots, \bm{A}^{(N)}}
\right]\kern-0.25em\right]
})\|_{\bm{\mathcal{W}}}^{2}.
\label{l}
\end{equation}

Our goal is to find matrices $\bm{A}^{(n)}$ for $n = 1, \cdots,N$ that minimize the equation (\ref{l}). The computation of the gradient of the fuction can be found in \cite{10}. After having the function and gradient, we can use any gradient-based optimization method \cite{11}  to handle this optimization problem.

We employed a first-order optimization approach, to be specific, the L-BFGS-B algorithm proposed by Richard H \textit{et al.}, to solve the weighted least squares problem. This algorithm functions as a gold stand- ard tool to solve large nonlinear optimization problems with simple bounds described. It develops a limited memory BFGS matrix to approach the Hessian of the objective function. The algorithm was devised to make good use of the form of the limited memory approximation to carry out the algorithm efficiently \cite{12}.

As mentioned previously, the rank of a tensor is often unknown, but results in \cite{13} showed that direct optimization methods have a better performance than alternating least square approaches when the rank is over- estimated. So bearing this fact in mind, we can set the number of compo nents $R$ relatively large in the beginning and check the results of this de composition. If the results are within our tolerance, we decrease R and check the results again. If they are still good enough, we tend to choose a relatively larger $R$.

Similar to CP decomposition, the results of CP-WOPT are a sequence
of rank-one tensors. Actually we can collect the vectors for each dimension and write the results in the form of factor matrices. If the tensor $\bm{\mathcal{X}}$ to be decomposed is of order 3, the results of CP-WOPT can be represented as  $\left[\kern-0.15em\left[
{\bm{A}, \bm{B}, \bm{C}}
\right]\kern-0.15em\right]$. If we pick up the $r$-th column vectors of these three matrices,
the vector outer product of these three vectors is the $r$-th rank-one tensor
of the results of the CP-WOPT decomposition. 
With these rank-one tensors, we are capable of reconstructing the original tensor. Let the sum of 
these rank-one tensors be \(\bm{\mathcal{X}}^{\prime}\). 
For each entry of \(\bm{\mathcal{X}}^{\prime}\), i.e. $x_{ijk}^{\prime}$ can written as the
sum:
\begin{equation*}
\sum_{r=1}^{R} a_{i r} b_{k r} c_{j r},
\end{equation*}
where $a_{ir}$ is the $i$-th element of the $r$-th column vector of $\bm{A}$, $b_{kr}$ is the $k$-th element of the $r$-th column vector of $\bm{B}$ and $c_{jr}$ is the $j$-th element of the $r$-th column vector of $\bm{C}$. 
For each element of $\bm{\mathcal{X}}$, $x_{ijk}$, no matter it is known
or unknown, there is en element of  \(\bm{\mathcal{X}}^{\prime}\),  $x_{ijk}^{\prime}$, corresponds to it. For the known entries, $x_{ijk}^{\prime}$
 and $x_{ijk}$ should be pretty close to each other, since it is the goal of our 
optimization. As a result, we can say that each $x_{ijk}$ 
corresponds to a sum of elements coming from the  factor matrices, $\bm{A}$, $\bm{B}$ and $\bm{C}$.

\subsubsection{Deep Neural Network}

The structure of the deep neural network (DNN) we employed in our DTF model is
a fully connected neural network with $D$ layers, where the $d$-th layer contains $U_{d}$ neurons. We used $\bm{r}_{d}^{n}$ to denote the input of the n-th sample into the $d$-th layer. Let $h(\cdot)$ be activation function, then the result of activation is $\bm{a}_{d}^{n} = h(\bm{r}_{d}^{n})$. Particularly, for the input layer, 
$\bm{a}_{0}^{n} = \bm{r}_{0}^{n}$.

Typically, we used forward propagation to calculate the input of next layer. For instance, for the input of ($d+1$)-th layer ($0 \leq  d \leq D$), $\bm{a}_{d+1}^{n}$ can be given by $
\bm{W}_{d} \bm{a}_{d}^{(n)}+\bm{b}_{d}$, where $\bm{W}_{d}$ is a $U_{d+1} \times U_{d}$ matrix and $\bm{b}_{d}$ is a bias. We optimized the parameters of the DNN’s $\bm{W}_{d}$ and $\bm{b}_{d}$ in order to minimize the loss function $F = 
\sum_{n} \operatorname{loss}\left(y^{\prime(n)}, y^{(n)}\right)
$, where $y^{\prime(n)}$ is  the predicted probability of the synergistic drug pairs and $y^{(n)}$
is the true synergy status.
% In our model, since we defined the prediction problem as a binary classification problem, we chose binary cross entropy  as the loss function.

We considered different hyperparameter settings for the DNN of DTF.   Besides different data normalization measures, different numbers of the layers and different hidden units inside a layer were studies. Further, we tried different learning rates and regularization techniques. The considered hyperparameter grid is showed in Table and more details will be given next.

We tested three different data normalization measures: (i) standardizing all features to zero mean and unit variance, (ii)standardization and applied hyperbolic tangent and (iii) standardization, hyperbolic tangent and standardization again. The situation of no data normalization was also considered. 
The hidden layer utilized relu activations, and the output layer used sigmoid activation. The binary cross entropy  was the loss function to be minimized. We tested two or three hidden layers, which are summarized in \textbf{Table \ref{T1}}.
We used Adam with learning rate $10^{-3}, 10^{-4}, 10^{-5}, 5 \times 10^{-3}, 5 \times 10^{-4},  5 \times 10^{-6} $ 
%0.00001, 0.000005 
as optimizer.
Early-stopping and dropout are considered as regularization 
techniques. For dropout, we considered dropout rate 0.1,0.2 or no drop out for input layer, and dropout rate 0.1, 0.2, 0.3, 0.4, 0.5 or no drop out for all hidden layers. We employed grid search to determine the best hyperparameters.

\begin{table}[htbp]
  \centering
  \caption{Hyperparameter settings considered for DNN of DTF}
  \vspace*{0.4em}
    \begin{tabular}{ll}
    \toprule
    Hyperparameter & Values Considered \\
    \midrule
    Preprocessing & no preprocessing; norm; norm + tanh; norm + tanh + norm \\
    Hidden units & [1024, 1024, 512], [2048, 2048, 1024], [2048, 1024, 512] \\
          & [512, 512, 512], [2048, 2048, 2048], [1024, 1024, 1024] \\
          & [2048, 2048], [1024,1024], [512, 512], \\
          & [2048, 1024], [1024, 512] \\
    Learning rates & $10^{-3}; 10^{-4}; 10^{-5}; 5 \times 10^{-3}; 5 \times 10^{-4};  5 \times 10^{-6} $  \\
    Dropout  & no dropout; input: 0.1, 0.2, hidden: 0.1, 0.2, 0.3, 0.4, 0.5 \\
    \bottomrule
    \end{tabular}%
     \caption*{\textit{Note:} All possible combinations of the
     hyperparameters in the table were optimized via grid-search.}
  \label{T1}%
%   \caption*{\textit{Note:} All possible combinations of the above hyperparameters in the table were optimized via grid-search.}
\end{table}%

\subsubsection{Feature Engineering}
 The key part of the DTF model is to generate features using the output of CP-WOPT, which is typically known as feature engineering. 
 
 As aforementioned, in the situation of three-dimensional tensor,
 the decomposition
results of CP-WOPT can be represented using factor matrices $\left[\kern-0.15em\left[
{\bm{A}, \bm{B}, \bm{C}}
\right]\kern-0.15em\right]$. Each matrix collects the latent information of a specific dimension. In particular, we can say that  $\bm{A}$ collects all the vectors corresponding to the latent information
of the first axis, $\bm{B}$ corresponding to the second axis, and $\bm{C}$ corresponding to the third axis. For each known  $x_{ijk}$, there is a sum, i.e.,  $\sum_{r=1}^{R} a_{i r} b_{k r} c_{j r}$, corresponding to it (\textbf{Figure 1. B})
as we have discussed in last subsection. Obviously, \(\left(a_{i 1}, \cdots, a_{i R}\right)\) is the $i$-th row vector of $\bm{A}$, which can be which can be regarded as features from the latent information of drug A. We applied the same principle to drug B and cell line. Finally, we got three feature vectors  \(\left(a_{i 1}, \cdots, a_{i R}\right)\),  \(\left(b_{k 1}, \cdots, b_{k R}\right)\) and  \(\left(c_{j 1}, \cdots, a_{j R}\right)\)  for a specific $x_{ijk}$. We collected them all together as the features of $x_{ijk}$. 
 
\subsubsection{Model Construction}
To build the DTF model to predict synergistic drug pairs, we need to link the CP-WOPT and DNN models together. And the bridge that connects CP-WOPT and DNN models is the feature enngineering described in last subsection. The overall construction of the DTF model and more details will be given below.

After After data collection and preprocessing, we built the three-order tensor with missing values  $\bm{\mathcal{X}}$ 
derived from original DDS data.
And the three dimensions of the tensor represent drug A, drug B and cell line, respectively. 
Note that the value for each element of $\bm{\mathcal{X}}$ is the original drug synergy score rather than the 0/1 labels which were generated by binarization. 
The values of the unknown entries of $\bm{\mathcal{X}}$ o not matter, since they were ignored during the computation process of CP-WOPT. In our model, we simply set them as 0, and we recorded the positions of the missing values in a position tensor $\bm{\mathcal{P}}$, which is of the same size as  $\bm{\mathcal{X}}$. A particular entry of $\bm{\mathcal{P}}$ is 1, if the corresponding element of $\bm{\mathcal{X}}$ is known, otherwise is 0.

Let R be the number of components, then we implemented CP-WOPT on $\bm{\mathcal{X}}$ (\textbf{Figure 1. B}), which required three parameters: the tensor $\bm{\mathcal{X}}$ to be 
decomposed, the position tensor 
 $\bm{\mathcal{P}}$ and the number of components $R$ we
wish to decompose the tensor into. As aforementioned, the decomposition
results of CP-WOPT can be represented using factor matrices  $\left[\kern-0.15em\left[
{\bm{A}, \bm{B}, \bm{C}}
\right]\kern-0.15em\right]$. 
Each matrix collects the latent information of a specific dimension. Here, it is resonable for us to assume that $\bm{A}$ collects all the vectors corresponding to the latent information
of drug A, $\bm{B}$ corresponding to drug B, and $\bm{C}$ corresponding to cell line. Based on the feature engineering mentioned above, we are able to get features for each synergy score $x_{ijk}$, which can be used to train the DNN.

If we set the number of components of CPWOPT as R, then for drug A, drug B and cell line, the number of dimension of each of them will be $R$, which suggests the parameters between the input layer of the DNN and the first layer is a matrix of size $U_{1} \times 3R$. The function TRAIN implemented the process of training the DNN, which used the features generated by the CP-WOPT and the labels from the binarization of synergy scores, namely, the synergy status. 
{$\bm{W}_{d}$} and {$\bm{b}_{d}$} are the parameters of the DNN, which defined the prediction process of the whole model. During the training process, we employed the classic forward propagation algorithm to calculate the results of the model being trained and the backward propagation algorithm to derive the gradient of the parameters of the DNN.

After training the DNN model, we used the trained model to predict the synergy status of drug pairs. For each unknown entry in the original tensor $\bm{\mathcal{X}}$, it is evident that there are also three feature vectors corresponding to  it. These features are denoted as $\bm{a}^{f}, \bm{b}^{f}, \bm{c}^{f}$, which 
represent drug A, drug B
and cell line features, respectively. And they can be input into the trained DNN model to predict the synergy status of any given drug pairs. This prediction process was implemented in the function PREDICT as shown in  \textbf{Algorithm \ref{dtf}}. We used forward propagation algorithm to calculate the prediction results.

The whole algorithm of our model is shown in the pseudocode of DTF (\textbf{Algorithm \ref{dtf}}).
The procedure MODEL incorporates all the functions to im- plement the prediction of synergy status of drug pairs, where $\bm{y}^{\prime}$ is a vector, which is used to collect all the predicted probability of the unknown entries. 
\vspace{1em}
\begin{breakablealgorithm}
\caption{DTF}
\begin{algorithmic}[1]
\State \textbf{Input:} tensor with missing values $\bm{\mathcal{X}}$, positions of missing values $\bm{\mathcal{P}}$,
 \State number of component $R$, synergy labels of drug pairs $\bm{l}$;
\State \textbf{Output:} predicted probability of missing pairs $\bm{y}^{\prime}$ ;
\Statex
% \State \textbf{Other Notations:} feature vectors $\bm{a,b,c}$, parameters of DNN $\left\{\bm{W}_{d}\right\}$,\\$\left\{\bm{b}_{d}\right\}$,
%  predicted probability of synergy status being 1 on particular 
% test sampel $y^{\prime}$,
%  factor matrices $\left[\kern-0.15em\left[ {\textbf{A,B,C}} \right]\kern-0.15em\right]$ 
% to serve as features;

% \Function{CPWOPT}{$\mathcal{X}$,$P$,$R$}: 
% %\Statex
% \State \textbf{Input:} tensor with missing values $\mathcal{X}$, positions of missing values $P$,
% \State number of component$R$;
% \State \textbf{Output:} factor matrices $\left[\kern-0.15em\left[ {\textbf{A,B,C}} 
%  \right]\kern-0.15em\right]$;
% %\Statex
% \State \textbf{return} $\left[\kern-0.15em\left[ {\textbf{A,B,C}} 
%  \right]\kern-0.15em\right]$; \Comment{The detailed implemention is omitted here}
% \EndFunction
\Function{PREDICT}{$\bm{a}^{f}$,$\bm{b}^{f}$,$\bm{c}^{f}$,$\left\{\bm{W}_{d}\right\}$,
$\left\{\bm{b}_{d}\right\}$}:
%\Comment Feature vectors $\bm{a}^{f},\bm{b}^{f},\bm{a}^{f}$
% \State \textbf{Input:} feature vectors $\bm{a,b,c}$, parameters of DNN $\left\{\bm{W}_{d}\right\}$,$\left\{\bm{b}_{d}\right\}$;
% \State \textbf{Output:} predicted synergy status of the test sample;
\State $y^{\prime}$ $\gets$ forwardprop($\bm{a}^{f}$,$\bm{b}^{f}$,$\bm{a}^{f}$,$\left\{\bm{W}_{d}\right\}$,
$\left\{\bm{b}_{d}\right\}$);
\State \textbf{return} $y^{\prime}$ 
\Comment Feature vectors $\bm{a}^{f}, \bm{b}^{f}, \bm{c}^{f}$
\EndFunction

\Statex
\Function{TRAIN}{$\left[\kern-0.15em\left[ {\bm{A,B,C}}
 \right]\kern-0.15em\right]$, $\bm{l}$}:
 % \State \textbf{Input:} factor matrices $\left[\kern-0.15em\left[ {\textbf{A,B,C}} 
 % \right]\kern-0.15em\right]$ serving as features, parameters of 
 % \State DNN $\left\{\bm{W}_{d}\right\}$,
 % $\left\{\bm{b}_{d}\right\}$;
 % \State \textbf{Output:} parameters of trained DNN $\left\{\bm{W}_{d}\right\}$,
 % $\left\{\bm{b}_{d}\right\}$;

\State $\left\{\bm{W}_{d}\right\} \gets$ init($\left\{\bm{W}_{d}\right\}$);
% \Comment Use Glorot uniform
\State $\left\{\bm{b}_{d}\right\} \gets \left\{\bm{0}\right\};$ 
% \qquad \qquad \qquad \qquad \qquad \qquad \qquad \qquad initializer
\For{$epoch$ $\gets$ 1  \algorithmicto\ $maxepoch$}
\State $\left\{\frac{\partial F}{\partial \bm{W}_{d}}\right\} \gets {0}$,
$\left\{\frac{\partial F}{\partial \bm{b}_{d}}\right\} \gets {0}$;

\For{\textbf{i} $\gets$ $mini\_batch\_indices$}
\State $\bm{y}^{\prime(i)}$ $\gets$  forwardprop($\bm{a}^{(i)}$,
$\bm{b}^{(i)}$,$\bm{c}^{(i)}$,$\left\{\bm{W}_{d}\right\}$,
$\left\{\bm{b}_{d}\right\}$)

\State $\left\{\frac{\partial F}{\partial \bm{W}_{d}}\right\},
\left\{\frac{\partial F}{\partial \bm{b}_{d}}\right\} \gets
\left\{\frac{\partial F}{\partial \bm{W}_{d}}\right\},
\left\{\frac{\partial F}{\partial \bm{b}_{d}}\right\}$  + backprop($\bm{a}^{(i)}$,
\State $\bm{b}^{(i)},\bm{c}^{(i)},\bm{y}^{(i)},\left\{\frac{\partial F}{\partial \bm{W}_{d}}\right\},
 \left\{\frac{\partial F}{\partial \bm{b}_{d}}\right\}$);
%  + backprop($\bm{a}^{(i)},
% \bm{b}^{(i)},\bm{c}^{(i)},\bm{y}^{(i)},\left\{\frac{\partial F}{\partial \bm{W}_{d}}\right\},
% \left\{\frac{\partial F}{\partial \bm{b}_{d}}\right\}$ )
\EndFor
\State $\left\{\bm{W}_{d}\right\}$,$\left\{\bm{b}_{d}\right\}$
 $\gets$ Adam($\left\{\bm{W}_{d}\right\}$,$\left\{\bm{b}_{d}\right\}$,$\left\{\frac{\partial F}{\partial \bm{W}_{d}}\right\},\left\{\frac{\partial F}{\partial \bm{b}_{d}}\right\}$);
\EndFor
\State \textbf{return} $\left\{\bm{W}_{d}\right\}$,$\left\{\bm{b}_{d}\right\}$;
\Comment The parameters of deep neural network
%$\left\{\bm{W}_{d}\right\} \gets$ init_glorot_uniform($\left\{\bm{W}_{d}\right\}$)
\EndFunction
\Statex

\Procedure{MODEL}{}:
%{$\mathcal{X}$,$P$,$R$}:\Comment{Implement the whole model}
% \State \textbf{Input:} tensor with missing values $\mathcal{X}$, positions of missing values $P$,
% \State number of component$R$;
% \State \textbf{Output:} total loss on the test set
\State $\left[\kern-0.15em\left[ {\bm{A,B,C}} 
 \right]\kern-0.15em\right]$ $\gets$ CP-WOPT(${\mathcal{X}}$,$\bm{\mathcal{P}}$,$R$);
 \Comment{Tensor factorization}
\State $\left\{\bm{W}_{d}\right\}$,$\left\{\bm{b}_{d}\right\}$ $\gets$ TRAIN($\left[\kern-0.15em\left[ {\bm{A,B,C}}
 \right]\kern-0.15em\right]$,$\left\{\bm{W}_{d}\right\}$,$\left\{\bm{b}_{d}\right\}$);
 %\State set $Loss$ to zero;
 \State $\bm{y}^{\prime}$ $\gets$ [ ]; 
 \Comment Vector to collect the predicting result
 \For{\textbf{i} $\gets$ $miss\_set\_indices$}
\State $y^{\prime(i)}$ $\gets$ PREDICT($\bm{a}^{(i)}, \bm{b}^{(i)}, \bm{c}^{(i)},\left\{\bm{W}_{d}\right\},\left\{\bm{b}_{d}\right\}$);
\State $\bm{y}^{\prime} \gets$  $\bm{y}^{\prime}$.append($y^{\prime(i)}$);
%\State 
%$Loss$ $\gets$ $Loss$ + loss($\bm{y}^{\prime(i)}$,$\bm{y}^{(i)}$)
 \EndFor
 \State \textbf{return} $\bm{y}^{\prime}$
 %\textbf{return} $Loss$
%\Comment{to serve as features}
\EndProcedure
\end{algorithmic}
\label{dtf}
\end{breakablealgorithm}

\vspace{2em}
\subsection{Model and Comparison Evaluation}

To evaluate whether the deep learning method, i.e., the DNN is able to do a better job at extracting the latent
information from the DDS data, we compare it with the CP-WOPT baseline model, a classical statistical models for for predicting synergy status of drug pairs based on the features generated by CP-WOPT and a state-of-the-art model proposed to do the the same task as DTF.
\noindent
\begin{itemize}[leftmargin=*]
    \item \textbf{CP-WOPT Baseline Classifier.} Generally speaking, CP- WOPT can be treated as a binary classifier alone. Recall that we can make use of the results to reconstruct the original tensor, and for each unknown synergy status, the sum corresponding to it can be regarded as a tensor score. Since a higher synergy score means a better synergistic effect of a given drug pair, we can treat a higher tensor score measured from the re- constructed tensor as stronger synergic effect of a given drug pair. Hence, for a given threshold of the tensor score, the CP-WOPT can be a binary classifier, which can served as the baseline model.
    \item \textbf{Logistic Regression (LR).} 
    Logistic regression model   was used to compare DTF with a classical non-linear method. The features input into the logistic regression model are the same as DNN, which were generated and extracted from the output of CP-WOPT.
    
    \item \textbf{DeepSynergy.} Proposed by Kristina Preuer \textit{et al.} in 2017, DeepSynergy was definitely a state-of-the art model to help identify novel synergistical drug pairs. DeepSynergy uses chemical and genomic information as input information, a normalization strategy to account for input data heterogeneity, and conical layers to model drug synergies \cite{8}. 
    We used the same input data as DeepSynergy. Note that the DeepSynergy utilized more datasets than the DTF.
    And the DeepSnergy is originally intended to solve the regression problem, to revise it into a classifier,  we chose 30 as the threshold to binarise the predicted synergy scores for consistency. And we used the DNN with the best architecture propsed by Kristina Preuer \textit{et al.} in this comparison. 
\end{itemize}
Due to the fact that all the models to be compared cannot discern the drug combinations AB, represented in order A-B or B-A, we double the sample. To benchmark the performance of DTF with other methods, we employed a particular stratified cross validation method, where the drug combinations were chosen to leave out in test sets \cite{8}. Under this circumstance, a given drug pair selected in the test set in a cell line is also included in the training set of all other cell lines. Note that since the DTF model is integrated, in the crooss validation process, different tensors were composed and each time  the features for a particular drug pair can be different.

\subsection{Software and Global Parameters}

CP-WOPT was implemented in Matlab tensor\_tool box version 3.1 and the Matlab wrapper of L-BFGC-B was written by Stephen Becker \cite{15}. We employed the Keras version 2.2.4 and scikit-learn version 0.21.3 to implement the DNN model and other machine learning models.

The tensor we constructed is of order 3, which has 3 axes, representing drug A, drug B and cell line. The dimension of drug A and drug B is 38, and that of cell line is 39. The number of components to be decomposed R was set to 1000. Hence, the dimension of the input features the DNN model is 3000. The threshold we choose to binarize drug synergy score is 30.

The numerical calculations in this paper have been partially
done on the supercomputing system in the Supercomputing Center of Wuhan University. Thanks  Supercomputing Center of Wuhan University for supporting the numerical calculations of this paper.

% The architecture of DeepSynergy was determined by the hyperpara- meter selection procedure, whose results are given in Supplementary Table S7. This procedure identified that tanh normalization, comprising first standardization and then a hyperbolic tangent fol- lowed by a second standardization, performed best. Furthermore, DeepSynergy has conic layers. A possible explanation for the fact that conic layers perform well, is their regularizing effect. The lower number of parameters available in the higher layers, which forces the model to generalize by constructing only the most important rep- resentations of chemical properties of the input compound combina- tion. Additionally, a large number of units in the first layer (8192) performed better. A smaller learning rate (10��5) and dropout regula- rization were also essential for learning performant networks. Overall, DeepSynergy has a conic architecture with two hidden layers having 8192 neurons in the first and 4096 in the second hid- den layer. It uses tanh input normalization, has a learning rate of 10��5, an input dropout rate of 0.2 and a hidden layer dropout rate of 0.5.
\section{Result}
\subsection{DNN Architecture}
The architecture of the DNN of DTF was determined via grid-search. This process showed that the second normalization strategy, standardization and applied hyperbolic tangent, performed best. Furthermore, the coinc layers, where the number of hidden units decreases half in each hidden layer,  have a better performance. A possible explanation for the fact that coinc layers are better, is their effect in regularizing. The smaller number of parameters in the higher layers pushes the model to be more general by extracting the most important latent information of the input features, generated by CP-WOPT. What's more, a coinc layers with more layers work well, namely, the one has three hiddern layers. A relateive small learning rate ($10^{-5}$) and dropout techonique were also critical for the DNN of the DTF to do a good job. In general, the DNN of DTF has a conic architecture with three hidden layers, having 2048 units in the first, 1024 in the second and 512 in the third layer. It employs standardization and applied hyperbolic tangent normalization method, has a learning rate of $10^{-5}$, and has dropout rates, 0.2, 0.5 for input and hidden layer, respectively.

% [2048, 1024, 512]
%norm_tanh 
% tmp_lr = 0.000010   
% tmp_input_dp = 0.2
% tmp_first_dp = 0.5
% tmp_second_dp = 0.5
%performed best.
\subsection{Method Comparison}
We used CP-WOPT algorithm to decompose the  tensors generated from the stratified cross validation method mentioned above. After the tensor factorization, we got features of drug A, drug B and cell line and also the tensor score of each drug pair. The factorized features from the CP-WOPT were used to build the classification models of DTF,  LR and CP-WOPT. 
For DeepSynergy, we considered the same test sets and used the best architecture propose in \cite{8}.

We chose metrics that are typical for classification task: area under the receiver operator characteristics curve (ROC AUC), area under the precision recall curve (PR AUC), accuracy (ACC), balanced accuracy (BACC), precision (PREC), sensitivity (TPR), and Cohen’s Kappa.  The threshold to binarilize the synergy score is the 90\% percentile \cite{8}, we chose in this paper. Therefore, in order to compute some of the metrics the tensor scores were binarilized using also the 90\% percentile of the tensor scores.
And we binarilized the predicting probability with threshold 0.5. 
The reuslts are summerized in \textbf{Table \ref{t2}}.
\begin{table}[htbp]
  \centering
  \caption{Methods comparision results based on performance metrics for the classification task.}
  \vspace*{0.4em}
    \begin{tabular}{lllllllll}
    \toprule
    Performance Metric & ROC AUC & PR AUC & ACC   & BACC  & PREC  & TPR      & Kappa \\
    \midrule
    DTF   & \textbf{\textit{0.89$\pm$0.02}}    & \textbf{\textit{0.57$\pm$0.04} }   & \textbf{0.93$\pm$0.01}    & \textbf{0.67$\pm$0.03}  & \textbf{0.73$\pm$0.03}    & \textbf{\textit{0.36$\pm$0.07} }   & \textbf{0.45$\pm$0.06} \\
    DeepSynergy &   \textbf{0.90$\pm$0.02} & \textbf{0.60$\pm$0.05} & \textbf{0.93$\pm$0.01}  & \textbf{0.67$\pm$0.03} & \textbf{\textit{0.72$\pm$0.04}} & \textbf{0.36$\pm$0.05} & \textbf{\textit{0.44$\pm$0.05}} \\
    Logistic Regression & 0.83$\pm$0.02 & 0.38$\pm$0.05 & 0.92$\pm$0.01 & 0.57$\pm$0.03 & 0.63$\pm$0.06 & 0.16$\pm$0.06 &  0.22$\pm$0.06 \\
    CP-WOPT  & 0.67$\pm$0.11 & 0.24$\pm$0.09 & 0.86$\pm$0.03 & 0.61$\pm$0.07 & 0.26$\pm$0.08 & 0.31$\pm$0.13  & 0.20$\pm$0.11 \\
    \bottomrule
    \end{tabular}%
  \label{t2}%
   \caption*{\textit{Note:} All values are average values$\pm$one standard deviation. We used bold  font to represent the best and performance values. And bold with itlatic font was used to represent the sencond best performance values. The performance metrics provided are area under ROC curve (ROC AUC), area under precision-recall curve (PR AUC), accuracy (ACC), balanced accuracy (BACC), precision (PREC), sensitivity (TPR) and Kappa.}
\end{table}%
Obviously, DeepSynergy had the best overall performance among all methods, achieveing a ROC AUC, PR AUC,  ACC, BACC, PREC, TPR,  and Kappa of 0.90, 0.60, 0.93, 0.67, 0.72, 0.36, and 0.44, respectively.  However, DTF exhibited a slightly inferior performance than DeepSynergy. The percision of DTF even had a higher performance  than DeepSynergy. And DTF achieved the same values as DeepSynergy in ACC and BACC metrics.
Needless to say, CP-WOPT baseline classifier model did the worst job here. And Logistic Regression model helped to improve the performance of CP-WOPT. But compared to DTF, the improvement was far behind. 

To sum up, DNN significantly improved  the performance of CP-WOPT. And more importantly,
with far less data sets, namely smaller size of the input data,
the DTF model, which combines DNN and CP-WOPT achieved a good performance almost the same as DeepSynergy in this comparison conditions. 

\subsection{Order independence}
\begin{figure}
    \centering
    \includegraphics[width = 9cm]{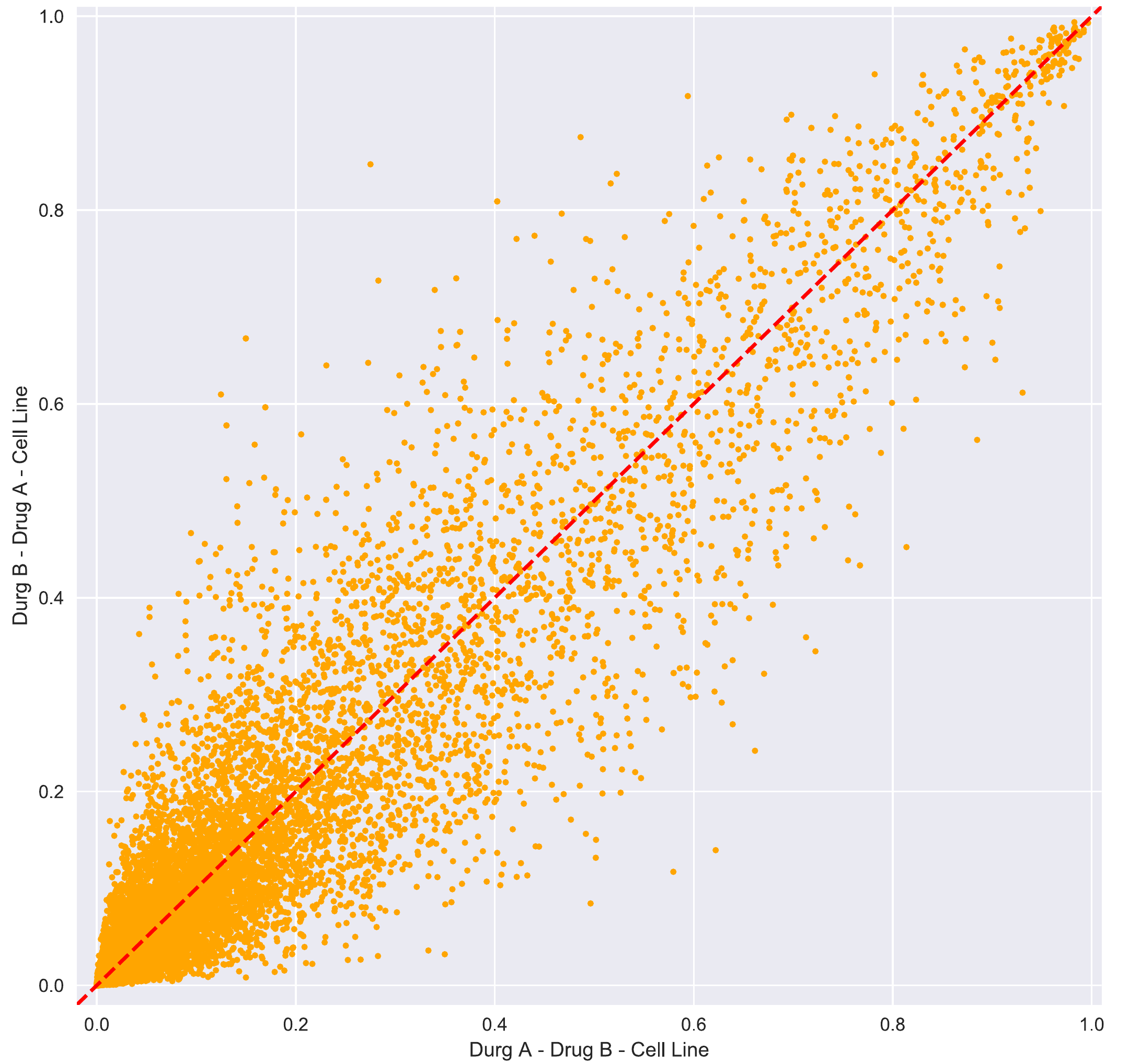}
    \caption{Scatter plot of the predictions derived from the two different orderings of drug pairs.  X-axis represents the ordering drug A - drug B - cell line, and Y-axis represents drug B - drug A - cell line.
    The Pearson correlation coefficient between the two predictions is 0.95.}
    \label{ordr}
\end{figure}

Drug combinations were doubled to DTF. We used both orders (drug A - drug B and drug B - drug A) for training and predicting. Undoubtedly, each pair was input into the DNN twice. The predicting results for the two different ways of ordering is shown in 
\textbf{Figure \ref{ordr}}.
Most of the values assemble around the original point, not too many points lied in the middle of [0,1] and a few gathered around the point (1,1), which is reasonable, since the data set is highly imbalanced after preprocessing. 
All points sat approximately two sides around the identity line, even though they are not quite close to it.
The predicting results of two different ordering achieved
a Pearson correlation coefficient 0.95, which exhibits that the DTF can ignore the order of the drug pair.

% Therefore, each combination was propagated twice through the network. Figure S1 shows the predictions for the two different ways of ordering. All values are close to the identity line and Pearson correlation coefficient of 0.98 was achieved, which shows that the network is able to neglect the order of the drug combination.

\subsection{Prediction of the Drug Pairs with Unknown Synergy Status}

The results above have shown that the DTF model we built is able to learn the relationship between the constructed tensor and the known synergy status of the drug pairs and the order of drug combinations can be neglected
, which means that we can reliably apply the model to predict the synergy status of the drug pairs we do not have synergy scores experimentally measured.To do this, we first constructed the tenor using approximately 90\% drug pairs with known synergy scores and then built the DTF-based prediction model based on the factorized features from the tensor. The about 10\% of the drug pair with known synergy scores were used for  final validation.

Using the same cutoff of 0.5 as we used for model evaluation, on validation set, DTF achieved,
ROC AUC, PR AUC,  ACC, BACC, PREC, TPR,  and Kappa of 0.90, 0.66, 0.93, 0.77, 0.66, 0.58, and 0.57, respectively. This ensured that our predictions on those drug combinations with synergy scores known were convincing. As a result, there 39 drug combinations out of 4680 drug combinations with unknown synergy scores are predicted with a probability  of highly synergistic higher than 0.5. The results are exhibited in \textbf{Table \ref{ll}}. Since here we what we predicted was the probability of highly synergistic drug pairs, it is reasonable to have relative small results with a probability above 0.5 to be highly synergistic.
    
\begin{center}
\begin{longtable}{lllll}
% Table generated by Excel2LaTeX from sheet 'final_0.5'
  \caption{Top predicted synergistic drug pairs.} %with a probability  of highly synergistic higher than 0.5.
   \\
    \toprule
    Cell line & Cancer & Drug A & Drug B & Probility \\
    \midrule
    CAOV3 & Ovarian & Dexamethasone & Etoposide & 0.99950922 \\
    CAOV3 & Ovarian & Cyclophosphamide & Etoposide & 0.99790645 \\
    CAOV3 & Ovarian & Etoposide & SN-38 & 0.99739087 \\
    CAOV3 & Ovarian & Carboplatin & Etoposide & 0.98608768 \\
    CAOV3 & Ovarian & 5-FU  & Etoposide & 0.98177201 \\
    CAOV3 & Ovarian & Etoposide & Metformin & 0.96313089 \\
    CAOV3 & Ovarian & Etoposide & Mitomycine & 0.94149995 \\
    OV90  & Ovarian & SN-38 & Vinorelbine & 0.92593026 \\
    CAOV3 & Ovarian & Etoposide & Topotecan & 0.90996122 \\
    OV90  & Ovarian & Topotecan & Vinorelbine & 0.90770292 \\
    CAOV3 & Ovarian & Etoposide & Vinblastine & 0.9032594 \\
    NCIH460 & Lung  & Etoposide & Paclitaxel & 0.89135432 \\
    HT144 & Melanoma & Etoposide & Paclitaxel & 0.87608039 \\
    A375  & Melanoma & Etoposide & Paclitaxel & 0.86257684 \\
    CAOV3 & Ovarian & Cyclophosphamide & Dexamethasone & 0.82120794 \\
    SKMES1 & Lung  & Dexamethasone & Paclitaxel & 0.80908144 \\
    CAOV3 & Ovarian & Doxorubicin & Etoposide & 0.79553449 \\
    HT144 & Melanoma & Etoposide & Topotecan & 0.79118788 \\
    CAOV3 & Ovarian & Etoposide & Paclitaxel & 0.77346706 \\
    CAOV3 & Ovarian & Etoposide & Oxaliplatin & 0.74256992 \\
    KPL1  & Breast & Carboplatin & Dexamethasone & 0.69261187 \\
    OV90  & Ovarian & Paclitaxel & Vinorelbine & 0.6880514 \\
    ES2   & Ovarian & Paclitaxel & Vinblastine & 0.6860581 \\
    OV90  & Ovarian & Etoposide & Vinorelbine & 0.67436397 \\
    KPL1  & Breast & Dexamethasone & Vinblastine & 0.67418796 \\
    VCAP  & Prostate & Etoposide & Topotecan & 0.66042769 \\
    OV90  & Ovarian & Mitomycine & Vinorelbine & 0.63128829 \\
    OV90  & Ovarian & Carboplatin & Vinorelbine & 0.63081199 \\
    OV90  & Ovarian & Vinblastine & Vinorelbine & 0.60110456 \\
    KPL1  & Breast & Dexamethasone & SN-38 & 0.59713888 \\
    CAOV3 & Ovarian & Etoposide & Vinorelbine & 0.58822578 \\
    SKOV3 & Ovarian & Dexamethasone & Vinblastine & 0.57798028 \\
    NCIH460 & Lung  & Paclitaxel & SN-38 & 0.55808449 \\
    CAOV3 & Ovarian & Dexamethasone & SN-38 & 0.55116999 \\
    VCAP  & Prostate & Etoposide & Vinblastine & 0.55102927 \\
    OV90  & Ovarian & Cyclophosphamide & Vinorelbine & 0.5375827 \\
    A375  & Melanoma & Etoposide & Vinorelbine & 0.53126353 \\
    VCAP  & Prostate & Topotecan & Vinblastine & 0.50386643 \\
    VCAP  & Prostate & SN-38 & Vinblastine & 0.50151283 \\
    \bottomrule
  \label{ll}%
\end{longtable}
\end{center}

\vspace{-4em}
\section{Discussion}
Our predicted  high probability of synergistic pairs are consistent with previous studies. Examples of these are as following.  Etoposide and dexamethasone, shows the highest predicted probability  of synergy among the 23 pairs in the ovarian cancer cell line CAOV3. Etoposide is a DNA topoisomerase II inhibitor and has been approved by FDA for treating testicular and lung cancers \cite{19}. Oral etoposide has demonstrated efficacy as an advanced treatment option for platinum-resistant ovarian cancer patients \cite{20} and also as a maintenance chemotherapy for advanced ovarian cancer patients to improve the survival outcomes \cite{21}. Dexamethasone is a steroid used to reduce inflammation. Dexamethasone is a type of steroid medication that has powerful anti-inflammatory and immunosuppressant and has been used in cancer treatment
\cite{22}. A recent study has confirmed the efficacy of the etoposide and dexamethasone combination therapy in hemophagocytic lymphohistiocytosi treatment \cite{23}. Therefore, combining dexamethasone and etoposide may improve the efficacy of etoposide as a single agent for the ovarian cancer treatment.

%To be continued...

Since the DTF model employed just a single data set, the conciseness of the model may be enhanced, however, the interpretability of the model and the results may be reduced.  In the future, we are interested in expanding the model to incorporate more information resources into the DTF model to improve the interpretability of DTF model. More effort will be put into investigating the other structures of the DNN for the sake of improving the performance of the entire model.

\section{Conclusion}
There are two key steps for the proposed DTF model, 1) decomposing the tensor with missing entries constructed from the original drug synergy data to generate features of drugs and cell lines using the CP-WOPT algorithm; and 2) training the DNN model using the factorized features together with the observed labels (synergetic status of the drug pairs) to predict the synergistic effect of the drug pairs with unknown synergetic scores. The DTF method, by linking the CP-WOPT and the DNN, used only a single data source but significantly improved the performance of CP-WOPT. In addition, the DTF model achieved 
 almost same prediction performance as the state-of-the-art model, DeepSynergy, using far fewer data sets, suggesting its potential as a valuable tool for predicting and optimizing synergistic drug pairs in \textit{silico} and thus guiding in \textit{vitro} and in 
\textit{vivo} discovery of rational combination therapies.

\section{Funding}
This research was supported in part by Canadian Breast Cancer Foundation, Natural Sciences and Engineering Research Council of Canada, Mitacs, University of Manitoba and China Scholarship Council.

\clearpage

\bibliographystyle{unsrt}  


\end{document}